\documentclass[twocolumn]{aa}
\usepackage{graphicx}
\usepackage{txfonts}
\begin{document}
   \title{High-resolution radio imaging of the most distant radio quasar SDSS J0836+0054}

   \author{S.~Frey\inst{1}
           \and
	   Z.~Paragi\inst{2}
	   \and
	   L.~Mosoni\inst{3}
	   \and
	   L.I.~Gurvits\inst{2}
          }
%%%   \offprints{S. Frey, \email{frey@sgo.fomi.hu}}

   \institute{F\"OMI Satellite Geodetic Observatory, P.O. Box 585, H-1592 Budapest, Hungary\\
              \email{frey@sgo.fomi.hu}
         \and
             Joint Institute for VLBI in Europe, Postbus 2, 7990 AA Dwingeloo, The Netherlands\\
              \email{zparagi@jive.nl, lgurvits@jive.nl}
	 \and
             Konkoly Observatory of the Hungarian Academy of Sciences, P.O. Box 67, H-1525 Budapest, Hungary\\
              \email{mosoni@konkoly.hu}
             }

   \date{Received March 10, 2005; accepted April 19, 2005}

   \abstract{
We observed SDSS J0836+0054, the most distant radio-loud quasar known at present ($z=5.774$) with the European Very Long Baseline Interferometry (VLBI) Network at the 5~GHz frequency on 2003 November 4. The source is compact at the milli-arcsecond (mas) angular scale with a flux density of 0.34~mJy. The observations with the Very Large Array (VLA) made on the consecutive day support a conclusion that the radio emission from SDSS J0836+0054 is essentially confined within the central 40~pc. Based on our phase-referenced VLBI observation, we obtained the astrometric position of the source accurate to $\sim 2$~mas.

   \keywords{techniques: interferometric --
             radio continuum: galaxies --
	     galaxies: active --
	     quasars: individual: SDSS J0836+0054 --
	     quasars: individual: J0836+0052 --
             cosmology: observations    }
   }

   \maketitle
%
%________________________________________________________________

\section{Introduction}

SDSSp J083643.85+005453.3 (J0836+0054 in short) is one of the highest
redshift quasars known to date discovered using the multicolor imaging data from
the Sloan Digital Sky Survey (SDSS, Fan et al.\ \cite{Fan01}) with a redshift $z=5.82$.
A more accurate redshift value of $z=5.774$ was determined using near-infrared spectroscopy by Stern et al.\ (\cite{Ster03}). The object is one of the twelve $z>5.7$ quasars found in the SDSS to date (Fan et al.\ \cite{Fan00,Fan01,Fan03,Fan04}). Among these, J0836+0054 is the only one identified with a previously known radio source with a flux density of $\sim 1$~mJy at 1.4~GHz. More recently, two other high-redshift ($z>6$) SDSS quasars (J1048+4637 and J1148+5221) were detected at 10~$\mu$Jy-level flux densities with sensitive 1.4-GHz continuum Very Large Array (VLA) imaging observations (Carilli et al.\ \cite{Cari04}).

Based on the Eddington argument, the mass of the central black hole in J0836+0054 is estimated as $M=4.8\times10^{9} M_{\sun}$, provided that its emission is not magnified by beaming or gravitational lensing (Fan et al.\ \cite{Fan01}). It is possible in principle that a significant fraction of the most distant quasars are gravitationally lensed and thus are magnified by a factor of ten or more (e.g. Wyithe \& Loeb\ \cite{Wyit02a,Wyit02b}). This would lead to overestimating their black hole masses by the same factor.

However, extensive search for gravitationally lensed multiply imaged quasars in the $z>5.7$ SDSS sample provided no evidence so far for strong lensing in any of the cases investigated. Analysing the flux distribution of the Ly$\alpha$ emission of J1030+0524, Haiman \& Cen (\cite{Haim02}) constrained its magnification to a factor of less than $\sim 5$. The first four $z>5.7$ SDSS quasars (Fan et al.\ \cite{Fan00,Fan01}), including J0836+0054, have been detected in X-rays (e.g. Brandt et al.\ \cite{Bran02}; Schwartz \cite{Schw02}). Their X-ray images are best fitted with single components. Richards et al.\ (\cite{Rich04}) conducted a snapshot imaging survey with the Hubble Space Telescope to look for possible lensed companions to the same four high-redshift SDSS quasars, with negative result. A deep optical study of the field around the quasar J1044$-$0125 with the Subaru Telescope revealed no secondary image but a faint foreground galaxy that may cause a factor of 2 magnification (Shioya et al.\ \cite{Shio02}). No magnification was found after a similar study for another two objects (Yamada et al.\ \cite{Yama03}).
The only radio-loud quasar in the sample, J0836+0054 has been detected and imaged at the angular scale of $\sim10$ milli-arcsecond (mas) with the European Very Long Baseline Interferometry (VLBI) Network (EVN) at the 1.6~GHz frequency (Frey et al.\ \cite{Frey03}). Most if not all of the radio emission is confined to a single compact object within $\sim 70$~pc. The quasar is not multiply imaged by gravitational lensing at the level of brightness ratio $\loa7$ within a $4\arcsec \times 4\arcsec$ field.

The absence of strong gravitational lensing in the $z>5.7$ SDSS quasar sample suggests that their central black hole mass estimates are not seriously biased. Assuming that these active galactic nuclei (AGN) are radiating at around their Eddington luminosity, and their emission is not significantly beamed, the corresponding masses are $\sim10^9 M_{\sun}$ (e.g. Fan et al.\ \cite{Fan01}). This is challenging for hierarchical structure formation models, and places limit on the slope of the quasar luminosity function at high redshift (e.g. Wyithe \cite{Wyit04}).

Indeed, there is growing evidence for supermassive black holes formed already as early as $<1$~Gyr after the big bang. The X-ray spectral properties of the highest-redshift radio-quiet quasars J1030+0524 and J1306+0356 are similar to those of low-redshift ones (Farrah et al.\ \cite{Farr04}; Schwartz \& Virani\ \cite{Schw04}). Near-infrared spectral observations probing the rest-frame ultraviolet emission of three objects (J0836+0054, J1030+0524 and J1044-0125) suggest no evolution of quasar metallicity out to $z \approx 6$: the Fe/Mg abundance ratio is near or above the solar value (Freudling et al.\ \cite{Freu03}).
Maiolino et al.\ (\cite{Maio04}) obtained the dust extinction curve for J1048+4637 and explained it by dust produced by Type-II supernova explosions. It is most likely that the formation of the central supermassive black holes and the AGN activity is preceded by intense star formation at the early evolutionary stages of this class of objects.

The compact radio emission in radio-loud AGN is known to originate from pc-scale jets in the close vicinity of the central supermassive black hole via incoherent synchrotron emission. The source J0836+0054 offers a unique possibility to study a radio-loud quasar at $z \approx 6$ and to compare its high-resolution structure with that of its lower-redshift counterparts. Our earlier 1.6-GHz VLBI observation (Frey et al.\ \cite{Frey03}) proved that the source had a compact, nearly unresolved structure at a $\sim 10$~mas angular scale. By means of the observations presented here, we aimed at imaging J0836+0054 at an even better angular resolution, obtaining radio spectral information on the mas scale, and improving the accuracy of the source astrometric position.

For calculating linear sizes and luminosities, we adopt a flat cosmological model ($\Omega_{\rm m}=0.3$, $\Omega_{\Lambda}=0.7$ and
$H_{\rm{0}}=65$~km~s$^{-1}$~Mpc$^{-1}$) throughout this paper. In this model, 1~mas
angular separation corresponds to a linear separation of 6.28~pc, and the age of the universe is 1~Gyr at the distance of J0836+0054.

%________________________________________________________________

\section{The EVN observation and data reduction}

The VLBI experiment took place on 2003 November 4, at the frequency of $\nu=5$~GHz.
Eight antennas of the EVN supplied very sensitive data at a recording rate of 512~Mbit~s$^{-1}$. The Effelsberg (Germany), Hartebeesthoek (South Africa), Medicina (Italy), Nanshan (China), Noto (Italy), Onsala (Sweden), Sheshan (China) and Westerbork (the Netherlands) radio telescopes produced useful data for a total of 7.5 hours. Dual circular polarization receivers were used where available. The data were recorded with the tape-based Mark~IV recording system at most of the stations. Effelsberg recorded with the Mark~5 disk-based system.
At this data rate we were able to recover 64~MHz bandwidth per polarization using 2-bit sampling, which resulted in superior baseline sensitivity.
The correlation took place at the EVN Data Processor at the Joint Institute for VLBI in Europe (JIVE), Dwingeloo, the Netherlands. A short 0.5-s integration time and 32 spectral channels were used in order to be able to image a large field of view, to look for possible gravitationally lensed radio components as far as $\sim20\arcsec$.

Due to the low (sub-mJy) flux density of J0836+0054 expected from the VLA data published earlier by Petric et al.\ (\cite{Petr03}), we employed the technique of phase-referencing. This involves regularly interleaving observations between the target source and a nearby, bright and compact reference source (e.g. Beasley \& Conway\ \cite{Beas95}). The delay, delay rate and phase solutions derived for the phase-reference calibrator were interpolated and applied for J0836+0054 within the target--reference cycle time of 5 minutes. The target source was observed for $\sim215$-s intervals in each cycle. The total observing time on J0836+0054 was 4~h, although certain antennas were not available for the whole period.

We choose J0836+0052 as the phase-reference calibrator, a compact quasar with a favourably small angular separation of $\sim5\arcmin$ from J0836+0054. The J2000 right ascension ($\alpha = 8^{\rm{h}}36^{\rm{m}}15\fs791287$) and declination ($\delta = +0{\degr}53{\arcmin}0\farcs00173$) of the reference source in the International Celestial Reference Frame (ICRF) are available from the Very Long Baseline Array (VLBA) Calibrator Survey\footnote{{\tt http://www.vlba.nrao.edu/astro/calib/index.shtml}} (Beasley et al.\ \cite{Beas02}). The uncertainty is 0.7~mas and 1.3~mas in $\alpha$ and $\delta$, respectively.

\begin{figure}
\centering
  \includegraphics[clip=,bb=40pt 160pt 600pt 665pt,width=69mm,
angle=0]{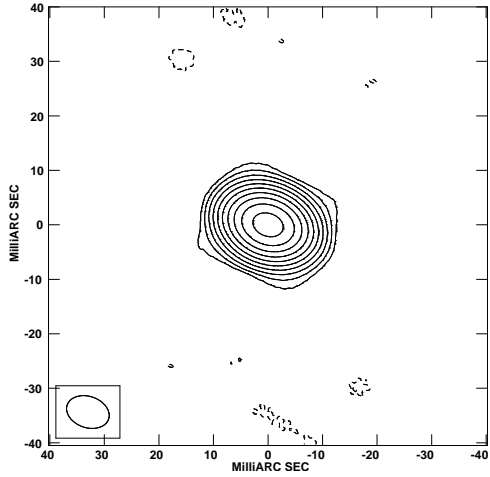}
  \caption{
A naturally weighted 5-GHz VLBI image of the phase-reference calibrator J0836+0052. The positive
contour levels increase by a factor of 2.
The first contours are drawn at $-0.3$ and 0.3~mJy/beam. The peak
brightness
is 218~mJy/beam. The restoring beam is 8.0~mas~$\times$~5.7~mas at a position
angle PA=$71{\degr}$ measured from north through east. The Gaussian restoring beam is indicated in the bottom left corner. The right ascension ({\it horizontal axis}) and declination ({\it vertical axis}) are related to the brightness peak.
   }
  \label{ref-5GHz}
\end{figure}

The US National Radio Astronomy Observatory (NRAO) Astronomical Image Processing System (AIPS, e.g. Diamond\ \cite{Diam95}) was used for the data calibration and imaging. The visibility amplitudes were calibrated using system temperatures measured at the antennas. Fringe-fitting was performed for the calibrator and fringe-finder sources (J0836+0052, J0927+3902 and J0839+0104) using 3-min solution intervals. The solutions were interpolated and applied to the data of J0836+0054. The AIPS tasks {\sc CALIB} and {\sc IMAGR} were used for the standard hybrid mapping procedure resulting in the naturally weighted image of the phase-reference calibrator J0836+0052 (Fig.~\ref{ref-5GHz}). The reference quasar shows a slightly resolved, compact radio structure. A uniformly weighted image not reproduced here was made with 2.4~mas~$\times$~1.0~mas restoring beam at PA=$12{\degr}$. It shows a single component as well, with a peak brightness of 134~mJy/beam.

\begin{figure}
\centering
  \includegraphics[clip=,bb=40pt 165pt 600pt 672pt,width=69mm,
angle=0]{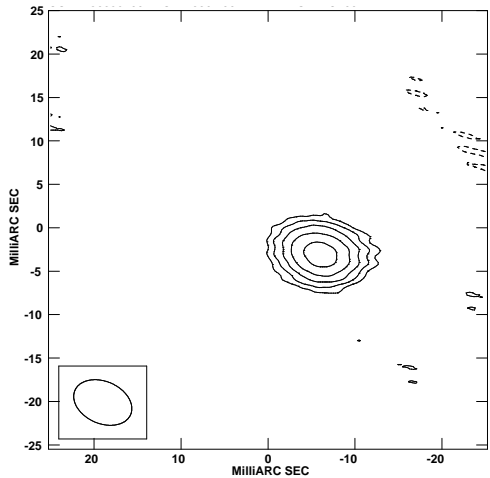}
  \caption{
A naturally weighted 5-GHz VLBI image of J0836+0054. The positive
contour levels increase by a factor of 2.
The first contours are drawn at $-70$ and 70~$\mu$Jy/beam. The peak
brightness
is 333~$\mu$Jy/beam. The restoring beam is 7.0~mas~$\times$~4.8~mas at PA=$66{\degr}$.
The coordinates are related to the a priori position derived from the 1.6-GHz VLBI experiment (Frey et al. \cite{Frey03}).
   }
  \label{target-5GHz}
\end{figure}

Before the imaging of the target source, the residual amplitude and phase corrections obtained after the imaging of the reference source using natural weighting were applied to the data of J0836+0054. We used natural weighting to minimise the image noise for a successful detection. The image made with {\sc IMAGR} after one cycle of {\sc CLEAN} iterations is shown in Fig.~\ref{target-5GHz}. The off-source rms image noise is $\sim35$~$\mu$Jy/beam, close to the expected thermal noise. The flux density of the source (the sum of the {\sc CLEAN} components) is $S=0.34$~mJy.

%------------------------Figure 1

%------------------------Figure 2

%________________________________________________________________

\section{The VLA observations and data reduction}

\begin{figure}
\centering
  \includegraphics[clip=,bb=40pt 165pt 600pt 668pt,width=69mm,
angle=0]{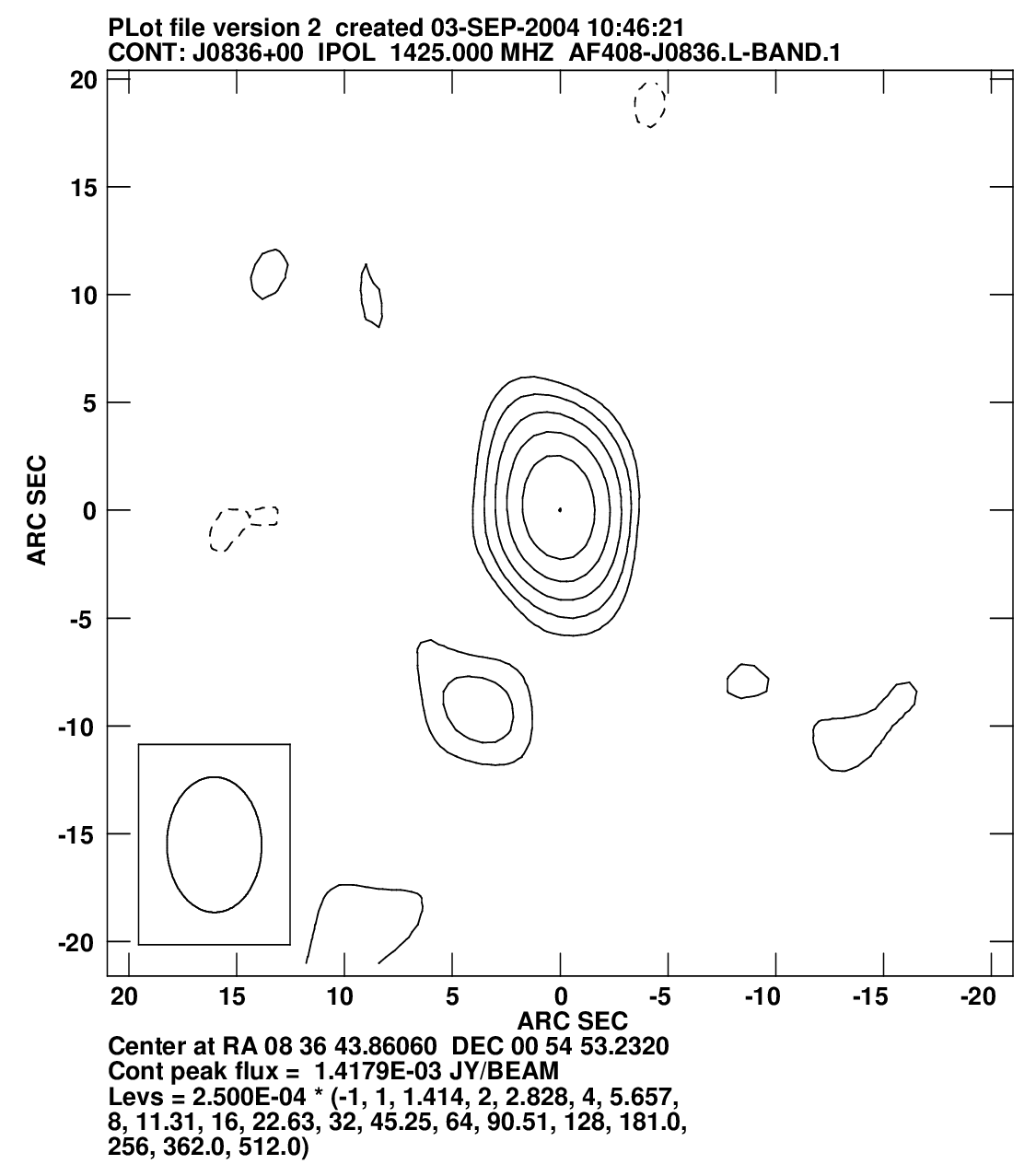}
  \caption{
A 1.6-GHz VLA image of J0836+0054. The positive
contour levels increase by a factor of $\sqrt2$.
The first contours are drawn at $-0.25$ and 0.25~mJy/beam. The peak
brightness
is 1.418~mJy/beam. The restoring beam is $6\farcs3 \times 4\farcs4$ at PA=$1{\degr}$.
   }
  \label{VLA-L}
\end{figure}

\begin{figure}
\centering
  \includegraphics[clip=,bb=40pt 165pt 600pt 668pt,width=69mm,
angle=0]{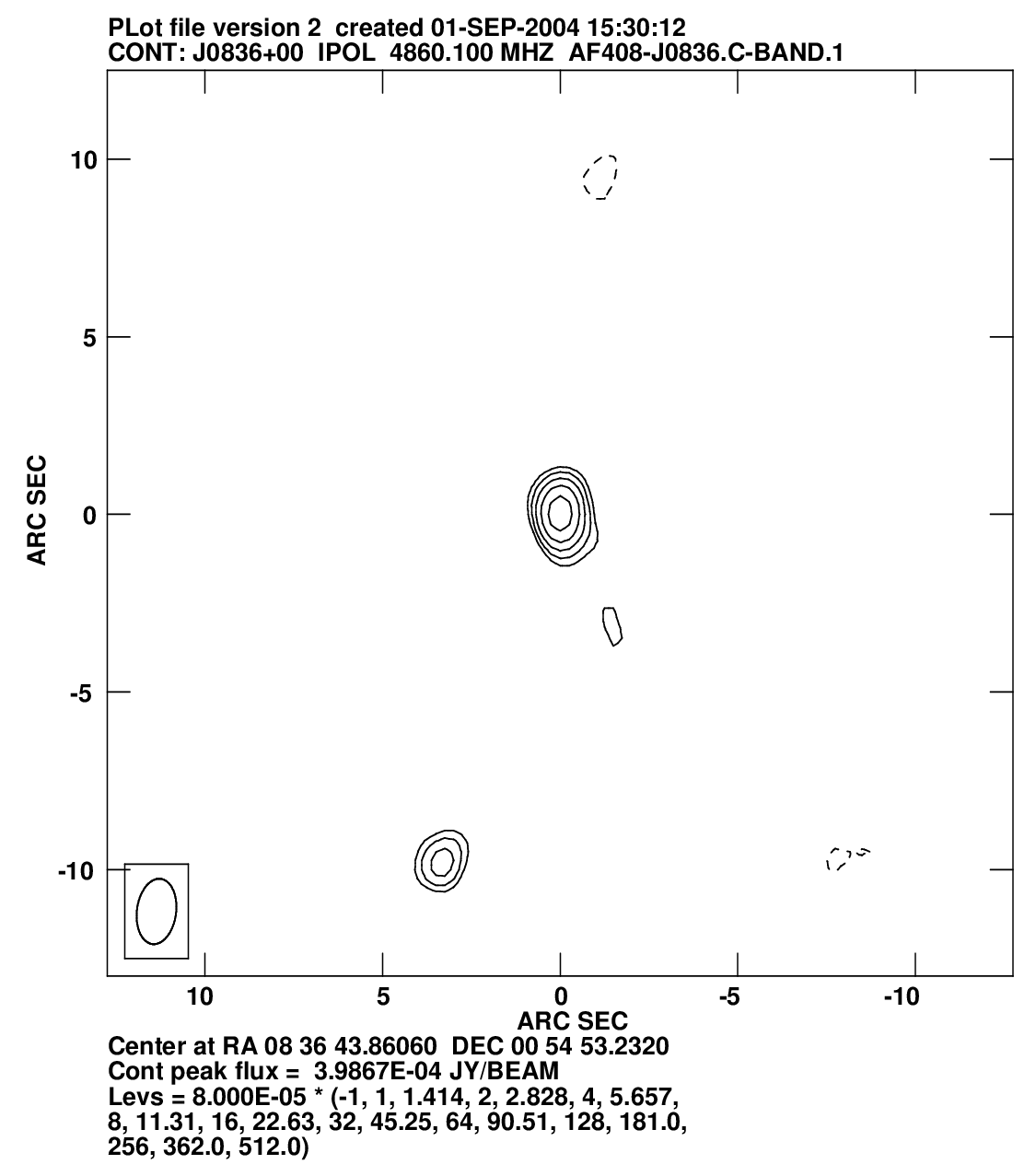}
  \caption{
A 5-GHz VLA image of J0836+0054. The positive
contour levels increase by a factor of $\sqrt2$.
The first contours are drawn at $-80$ and 80~$\mu$Jy/beam. The peak
brightness
is 399~$\mu$Jy/beam. The restoring beam is $1\farcs9 \times 1\farcs1$ at PA=$-7{\degr}$.
   }
  \label{VLA-C}
\end{figure}

Sensitive dual-frequency VLA observations of J0836+0054 were made earlier by Petric et al.\ (\cite{Petr03}), indicating a steep radio spectrum between 1.4 and 5~GHz (spectral index $\alpha_{\rm{sp}}=-0.8$; $S\propto\nu^{\alpha_{\rm{sp}}}$). Note that these observing frequencies correspond to 9.5~GHz and 33.9~GHz in the quasar rest frame due to the extremely high redhift. In order to assess the fraction of the flux density originating from the mas-scale compact core at 5~GHz, we conducted short supporting VLA observations of J0836+0054. The VLA data were taken on 2003 November 5, the day following the 5-GHz EVN experiment. The array was in its B configuration. The source was observed for 18.5 and 19.5 minutes at the 1.4 and 5-GHz bands, respectively, with a bandwidth of 50~MHz in both circular polarizations. Phases and amplitudes were calibrated in the standard way in AIPS. The source 3C~147 (total flux densities 21.4501~Jy at 1.4649~GHz and 7.9134~Jy at 4.8851~GHz) was used to define the absolute flux density scale. Self-calibration was performed with the sources in the field of view.  Our VLA images at 1.4 and 5~GHz centered on J0836+0054 are shown in Figs.~\ref{VLA-L} and \ref{VLA-C}. The flux densities measured ($1.96\pm0.31$~mJy and $0.43\pm0.06$~mJy at 1.4 and 5~GHz, respectively) are consistent with the values published by Petric et al. (\cite{Petr03}). Notably, the 5-GHz VLA flux density is close to the EVN value obtained one day earlier. This indicates that the quasar is indeed very compact: its radio emission comes from within the region imaged with VLBI.

%------------------------Figure 3

%------------------------Figure 4

%________________________________________________________________

\section{Results and discussion}

The equatorial coordinates determined for J0836+0054 are
$\alpha_{\rm{J2000}} = 8^{\rm{h}}36^{\rm{m}}43\fs86020$ and
$\delta_{\rm{J2000}} = +0{\degr}54{\arcmin}53\farcs2290$. The estimated positional uncertainty is 2~mas. Due to the small target--reference angular separation, the error in the absolute ICRF position of J0836+0054 is dominated by the uncertainty in the phase-reference calibrator source coordinates (cf. Chatterjee et al.\ \cite{Chat04}). The improved coordinates are consistent with the position derived from the 1.6-GHz VLBI data (Frey et al.\ \cite{Frey03}) within the errors.
Note that a different phase-reference calibrator source (J0839+0104) was used for the 1.6-GHz experiment. Its complex structure made the relative astrometry of J0836+0054 more uncertain in that observation.

A weak nearby radio source located at $9\farcs7$ to the south and $3\farcs4$ to the east of J0836+0054 reported earlier by Petric et al.\ (\cite{Petr03}) is seen in both VLA images (Figs.~\ref{VLA-L}, \ref{VLA-C}).
This secondary source has been investigated as a possible gravitationally lensed image of the extremely distant quasar. The lensing hypothesis has been ruled out on the basis of different radio spectral indices and a deep optical image (D. Rusin \& B. McLeod, private communication). The latter shows the second source significantly resolved, suggesting that it is associated with a lower-redshift galaxy. We are not aware of any optical spectroscopic observation that would directly prove that the secondary object is in the foreground. In this case, due to its large angular separation, it is very unlikely that the galaxy gravitationally magnifies the quasar emission by any significant factor.

With our 5-GHz VLBI data, we were able to check within the undistorted field of view whether the nearby $\sim200$~$\mu$Jy source present in the VLA image shows compact radio emission characteristic to AGN. Nothing was detected at the expected position at the minimum brightness level of $\sim160$~$\mu$Jy/beam (5~$\sigma$). Given the limited sensitivity, this however does not completely rule out the existence of a weak compact component in the secondary radio source.

The 5-GHz VLBI image (Fig.~\ref{target-5GHz}) indicates that J0836+0054 is compact at mas angular scales. The structure is reminiscent of a typical compact radio ``core'' often observed with VLBI in radio-loud AGN. Since the flux density in this core is practically identical with that measured with the VLA at the same time, the radio emission originates from the central region within $\sim 40$~pc linear extent.

For J0836+0054, the monochromatic radio luminosity at 5~GHz (rest frame) is $L_5=1.1\times10^{25}$~W~Hz$^{-1}$~sr$^{-1}$. The rest-frame 2--10~keV luminosity obtained from the Chandra X-ray observation is $L_{\rm X}=2.3 \times 10^{38}$~W (Schwartz \cite{Schw02}).
These values and the mass estimate in principle define the place of our object with respect to the ``fundamental plane'' of black hole activity derived recently by Merloni et al.\ (\cite{Merl03}) after studying correlations between the black hole mass, and the radio and X-ray luminosities of relatively nearby AGN as well as galactic stellar-mass black holes.
The quasar J0836+0054 is naturally located at the high-luminosity end of the relation. Following Merloni et al.\ (\cite{Merl03}), we calculate the 5-GHz radio luminosity as $L_{\rm R}=2\pi \nu L_5=3.5 \times 10^{33}$~W. Note that the integration over $2\pi$ solid angle most likely leads to an overestimate of the luminosity because the radio emission is possibly not isotropic but relativistically beamed in our case. Still, J0836+0054 lies very close to the best-fitting ``fundamental plane'' in the (log$L_{\rm R}$, log$L_{\rm X}$, log$M$) space, well within the substantial scatter.

%________________________________________________________________

\section{Conclusions}

J0836+0054, the most distant radio-loud quasar known to date ($z=5.774$) is  detected with EVN observations at 5~GHz as a compact source with a flux density of 0.34~mJy. The absolute ICRF position of the quasar is determined with the accuracy of 2~mas. Nearly contemporaneous VLA observations confirm its steep radio spectrum in the rest-frame frequency range $\sim10-30$~GHz, and imply that the radio emission is confined within the central 40~pc of the object. Based on its observed  mas-resolution structure and radio luminosity, J0836+0054 appears similar to other powerful radio-loud AGN observed at much later cosmological epochs. Our results provide further evidence for the existence of ``typical'' AGN powered by accretion onto supermassive black holes already at $z\approx6$.

%________________________________________________________________

\begin{acknowledgements}
We thank David Rusin and Brian McLeod for informing us about their optical observations of J0836+0054, and Andy Biggs for useful discussions.
This research was supported by the Hungarian Scientific Research Fund (OTKA, grant T046097) and the European Commission's FP6 I3 Programme RadioNet under contract No.\ 505818. SF acknowledges the Bolyai Research Scholarship received from the Hungarian Academy of Sciences.
The European VLBI Network is a joint facility of European, Chinese,
South African and other radio astronomy institutes funded by their
national research councils.
The National Radio Astronomy Observatory is a facility of the National Science Foundation opearted under cooperative agreement by Associated Universities, Inc.
This research has made use of the NASA/IPAC Extragalactic Database (NED) which
is operated by the Jet Propulsion Laboratory, California Institute of
Technology, under contract with the National Aeronautics and Space
Administration.
\end{acknowledgements}

%________________________________________________________________

\end{document}